\documentclass[11pt,a4paper]{article}
\usepackage{jheppub}
\usepackage{epsf}
\usepackage{amsmath}
\usepackage{amssymb}
\usepackage{amsfonts}
\usepackage{graphicx}
\usepackage{graphics}
\usepackage{ulem}
\usepackage{epstopdf}

\newcommand{\xt}{x_{_\perp}}
\newcommand{\pt}{p_{_T}}

\newcommand{\sqrtsnn}{\sqrt{s_{_{\mathrm{NN}}}}}

\newcommand{\A}{{\rm A}}

\newcommand{\rag}{R^{^\A}_{_G}}
\newcommand{\rafd}{R^{^\A}_{_{F_2}}}

\newcommand{\rpa}{R_{_{p\A}}}

\newcommand{\rdau}{R_{_{d {\rm Au}}}}
\newcommand{\rauau}{R_{_{{\rm Au Au}}}}

\def\cO#1{{{\cal{O}}}\left(#1\right)} 

\title{Inclusive prompt photon production in nuclear collisions at RHIC and LHC}

\author[a]{Fran\c{c}ois Arleo,}
\author[b,c]{Kari J. Eskola,}
\author[d]{Hannu Paukkunen,}
\author[d]{Carlos A. Salgado}

\affiliation[a]{Laboratoire d'Annecy-le-Vieux de Physique Th\'eorique (LAPTH)\\ UMR5108, Universit\'e de Savoie, CNRS, BP 110, 74941 Annecy-le-Vieux cedex, France}
\affiliation[b]{Department of Physics, P.O. Box 35, FI-40014 University of Jyv\"askyl\"a, Finland}
\affiliation[c]{Helsinki Institute of Physics, 
P.O. Box 64, FIN-00014 University of Helsinki, Finland}
\affiliation[d]{Departamento de F\'\i sica de Part\'\i culas and IGFAE, Universidade de Santiago de Compostela, Spain}

\emailAdd{arleo@lapp.in2p3.fr}
\emailAdd{kari.eskola@phys.jyu.fi}
\emailAdd{hannu.paukkunen@usc.es}
\emailAdd{carlos.salgado@usc.es}

\abstract{
Nuclear modification factors of inclusive prompt photon production  in $d$--Au collisions at RHIC and $p$--Pb collisions at the LHC are provided at different rapidities. The calculations are performed at 
 NLO accuracy using the EPS09 NLO nuclear parton distribution functions (nPDFs) and their error sets. The results are compared to the ones obtained with the nDS and HKN07 NLO nPDFs, and to the corresponding nuclear modification factors of neutral pion production in these collisions. 
 The sensitivity of these results to the scale choice is also investigated. Interestingly, the predictions using the different nPDF sets differ from each other to the extent that this observable can be expected to become very useful for probing 
 nPDFs
 over a wide range of Bjorken-$x$. 
In order to obtain a perturbative QCD baseline
in heavy-ion collisions,
calculations are 
 carried out for minimum bias Au--Au collisions at RHIC and Pb--Pb collisions at the LHC. We also estimate the maximal possible suppression which the produced QCD matter can be expected to have on inclusive prompt photon production
due to the quenching of the fragmentation component.
 The nuclear modification factor for prompt photon production is thus suggested to be used for gauging both the cold and the hot nuclear matter effects on other hard processes which are expected to be affected by quark-gluon plasma formation, such as  large-$\pt$ hadron and jet production.
}

\arxivnumber{xxxx.yyyy}

\begin{document} 
\maketitle
\setcounter{footnote}{0}
\renewcommand{\thefootnote}{\arabic{footnote}} 	

\section{Introduction}

The production of inclusive and isolated prompt photons in hadronic collisions is a process
that is known -- along with jet production -- to carry direct information on the gluon Parton Distributions Functions (PDF) in the proton~\cite{Aurenche:1988vi,Martin:1998sq}. However, prompt photon production data are presently not used in the global fits of PDFs but  rather provide an independent test of perturbative QCD and of the universality of the partonic densities. The good overall agreement between next-to-leading order (NLO) QCD calculations~\cite{Aurenche:2006vj} and world data from fixed-target ($\sqrt{s}=23$~GeV) to collider experiments ($\sqrt{s}=1.8$~TeV) indicates that the underlying parton dynamics is properly understood.
In this respect, the expected high-precision measurements to be performed at the LHC and RHIC might give extra constraints on the PDFs in the gluon sector~\cite{Ichou:2010wc}. The agreement between the first ATLAS and CMS measurements~\cite{Collaboration:2010sp,Collaboration:2010fm} and NLO calculations is thus one more encouraging step towards this goal.

The recent global analyses of nuclear PDFs (nPDF)~\cite{Eskola:1998df,Hirai:2004wq,Hirai:2001np,deFlorian:2003qf,Hirai:2007sx,Eskola:2008ca,Eskola:2009uj,Schienbein:2009kk} have shown that nuclear collisions seem to follow the same 
collinear factorization theorem that works in hadronic collisions:
Structure function data in nuclear (lepton) deep inelastic scattering (DIS) as well as Drell-Yan dilepton (DY) and hadron production measurements in high-energy proton(deuterium)-nucleus ($p(d)$--$A$) collisions
are consistent with universal (process independent) nuclear modifications of PDFs \cite{Eskola:2008ca,Eskola:2009uj}. Other checks have been carried out recently using neutrino-DIS data~\cite{Paukkunen:2010hb,Kovarik:2010uv}.
Whether such a good description persists for prompt photon production has however not been verified yet, as no decisive-precision $p$--$A$ or $d$--$A$ data are presently available. Keeping in mind the ongoing efforts at RHIC and appreciating the potential
capabilities of the LHC, this situation 
is expected to
undergo a change in the near future.

As discussed in~\cite{Arleo:2007js,BrennerMariotto:2008st}, prompt photon production in $p$--$A$ collisions appears to be a promising tool for probing the gluon nuclear densities, which up to now have 
been constrained through scaling violations of the nuclear DIS and DY data and inclusive $\pi^0$ production in $d$--Au collisions at RHIC \cite{Eskola:2009uj} but, clearly, for which more constraints would be badly needed. 
Therefore, complementary to recent studies of inclusive pion production ~\cite{QuirogaArias:2010wh} and prompt photon+heavy quark production~\cite{Stavreva:2010mw}, we present and discuss in the present paper the NLO predictions for the nuclear 
modification
ratios of prompt photon spectra in $d$--Au collisions at RHIC and $p$--Pb collisions at the LHC using modern nPDF sets (section~\ref{sec:pA}). Baseline predictions in heavy-ion ($A$--$A$) collisions will also be discussed in detail in section~\ref{sec:AA}. Before this, let us first briefly look at the theoretical framework used in the present analysis.

\section{Theoretical framework}\label{sec:framework}

\subsection{Elements of the calculation}
\label{sec:ingredients}

We consider in this analysis inclusive production of prompt photons,
\begin{equation}
 h_1\ +\ h_2 \to\ \gamma\ + {\rm X}, \nonumber
\end{equation}
 in $d$--Au and Au--Au collisions at RHIC
($\sqrtsnn=200$~GeV) as well as in $p$--Pb and Pb--Pb collisions at the LHC ($\sqrtsnn=8.8$~TeV and $\sqrtsnn=5.5$~TeV, respectively) nominal energies.

Using collinear factorization, the inclusive photon production cross section can be written as a convolution 
\begin{equation}
d\sigma_{h_1h_2\rightarrow \gamma+X} = \sum_{i,j} f_i^{h_1}(M^2)\otimes f_j^{h_2}(M^2)\otimes d\hat \sigma_{ij\rightarrow \gamma+X'}(\mu^2, M^2)+ {\cal O}(1/M^2),
\label{eq:factorization1}
\end{equation}
where $\otimes$ is an integral over the longitudinal momentum fractions $x_1$ and $x_2$ of the incoming partons $i$ and $j$, whose number densities - the PDFs of the projectile and target PDFs -- are $f_i^{h_1}(x_1,M^2)$ and $f_j^{h_2}(x_2,M^2)$. The factorization scale entering the PDFs and the perturbatively calculable partonic pieces (subcross-sections at LO) $d\hat\sigma_{ij\rightarrow \gamma+X'}$ is  denoted by $M$, and the  renormalization scale by $\mu$. As usual, $X$ indicates the inclusive nature of the cross section and $X'$ indicates that in each partonic hard process (2-to-2 or 2-to-3) we integrate over everything else but the photon.
 
At leading order (LO) in pQCD, $\mathcal O(\alpha_{\rm em}\alpha_s)$, photons can be produced directly via two types of partonic subprocesses: 
Compton scattering $q(\overline q) g \rightarrow \gamma q(\overline q)$ and 
annihilation $q\overline q \rightarrow \gamma g$. 
In terms of the transverse momentum $\pt$ and rapidity $y$ of the photon, and the cms-energy $\sqrt s$, the momentum fractions typically probed by direct photon production are $x_{1,2}\approx(2\pt/\sqrt s)e^{\pm y}$. Thus, 
towards smaller $\pt$, larger $y$ and larger $\sqrt s$, the process becomes sensitive to PDFs at smaller $x_2$ (target) and larger $x_1$ (projectile). At small values of $x$, the gluon distribution is larger than that of sea quarks, which is why the $q\bar{q}$ annihilation channel represents a small contribution to the cross sections in the entire kinematical domain reached at RHIC and LHC, see e.g.~\cite{Ichou:2010wc}. As a consequence, direct photon production is very sensitive to the gluon content of protons and nuclei at these colliders~\cite{Arleo:2007js}.

At next-to-leading order (NLO), $\mathcal O(\alpha_{\rm em}\alpha_s^2)$, there are altogether 7 types of contributing direct subprocesses (see~\cite{Aurenche:1987fs}): 
$q(\overline q) g \rightarrow \gamma q(\overline q)g$,
$q_i \overline{q_i} \rightarrow \gamma q_i\overline{q_i}$,
$q_i \overline{q_i} \rightarrow \gamma q_{j\ne i}\overline{q_j}$,
$q_i \overline{q_i} \rightarrow \gamma gg$,
$q_i(\overline{q_i}) q_{j\ne i}(\overline{q_{j}}) \rightarrow \gamma q_{i}(\overline{q_i}) q_j(\overline{q_j})$,
$q_i (\overline{q_i}) q_i(\overline{q_i}) \rightarrow \gamma q_i (\overline{q_i}) q_i (\overline{q_i})$, and
$gg \rightarrow \gamma q_i \overline{q_{i}}$. At LHC, $qg$ Compton scattering as well as gluon fusion are the processes which dominate the total cross section due to the high gluon density in the proton at small values of Bjorken-$x$.

In addition to the direct production channels discussed above, prompt photons can also be emitted through collinear fragmentation from high-$\pt$ quarks or gluons which are produced in primary hard partonic collisions. This is, again schematically, expressible as 
\begin{equation}
d\sigma_{h_1h_2\rightarrow \gamma+X} = \sum_{i,j} f_i^{h_1}(M^2)\otimes f_j^{h_2}(M^2)\otimes d\hat \sigma_{ij\rightarrow k+X'}(\mu^2, M^2,M_F^2) \otimes D_{\gamma/k}(M_F^2).
\label{eq:factorization2}
\end{equation}
The perturbatively calculable pieces related to the partonic hard processes, 
$d\hat \sigma_{ij\rightarrow k+X'}$, are now of the order $\mathcal O(\alpha_s^{2+n})$ ($n=0$ for LO, and 1 for NLO parton production). The collinear divergences associated to this source of photons are resummed and absorbed into scale-dependent fragmentation functions (FFs) into photons, $D_{\gamma/k}(z,M_F^2)$, where $z$ is the fractional momentum over which the last convolution is taken, and $M_F$ is the factorization scale related to the fragmentation process. The FFs scale asymptotically like $\mathcal O(\alpha_{\rm em}/ \alpha_s)$~\cite{Owens:1986mp,Aurenche:1998gv}, and thus the convolution of the FFs with the (higher-order) partonic cross sections makes the fragmentation contributions of the same order as the direct channel.

We present here results from a complete NLO calculation, which includes both the direct and the fragmentation processes at this order. The inclusive calculations performed here are based on the \texttt{INCNLO}-package~\cite{Aurenche:1998gv,Aurenche:1999nz}. We use the CTEQ6.6M NLO PDF sets for the free proton PDFs~\cite{Nadolsky:2008zw} and the Bourhis--Fontannaz--Guillet fragmentation functions of photons~\cite{Bourhis:1997yu}.
The renormalization and factorization scales we take to be equal and of the order of the photon transverse momentum, $\mu^2=M^2=M_F^2 = a\times\pt$. To investigate the uncertainty related to the perturbative nature of the calculation presented, we however vary the constant $a$ between $1/2$ and $2$.  
Finally, the prompt photon results are also contrasted to what is expected 
for inclusive single-pion production, 
using the AKK08 set~\cite{Albino:2008fy} of fragmentation functions.

\subsection{Nuclear Parton Distribution Functions}
\label{sec:npdf}

We estimate the nuclear modifications of the photon spectra
at RHIC and LHC using the 
recent NLO releases of nuclear PDFs:~nDS~\cite{deFlorian:2003qf}, EPS09~\cite{Eskola:2009uj} and HKN07~\cite{Hirai:2007sx}. All these sets quantify the scale-dependent ratios between the PDF of a proton inside a nucleus, $f_i^{p/A}$, and that in the unbound proton, $f_i^{p}$,
\begin{equation}\label{eq:npdfratio}
R_i^A(x,Q^2) \equiv \frac{f_i^{p/A}(x, Q^2)}{f_i^p(x, Q^2)}.
\end{equation}
These modifications are known to exhibit a rich structure as a function of $x$. The suppression at small $x\lesssim 10^{-2}$ is commonly referred to as shadowing while an enhancement (anti-shadowing) is predicted around $x\sim 10^{-1}$. At 
$x\gtrsim0.3$ the ratio becomes again smaller than 1 (EMC-effect), and larger than 1 just below $x=1$  because of the Fermi-motion in nuclei.

\begin{figure}[h]
    \begin{center}
      \includegraphics[height=7.5cm]{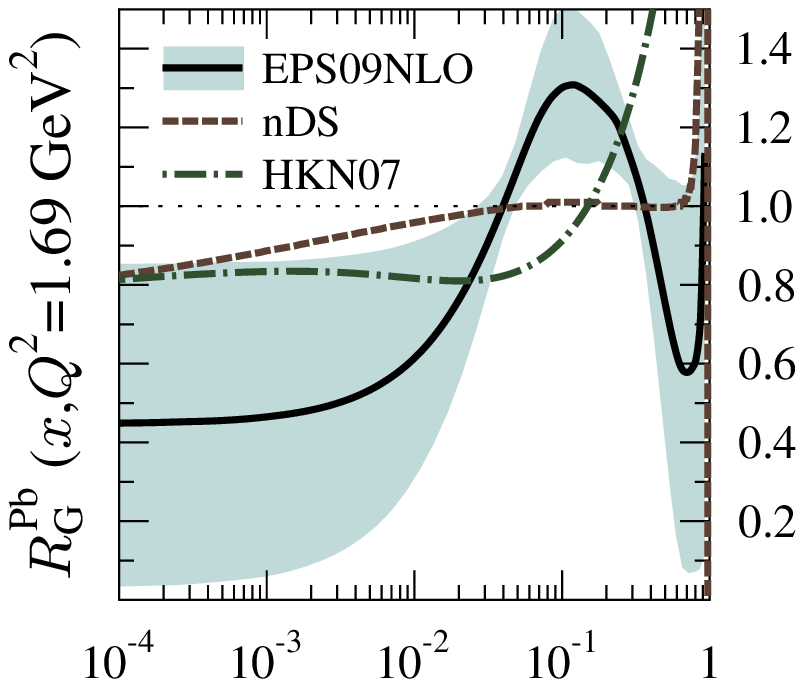}
      \includegraphics[height=7.5cm]{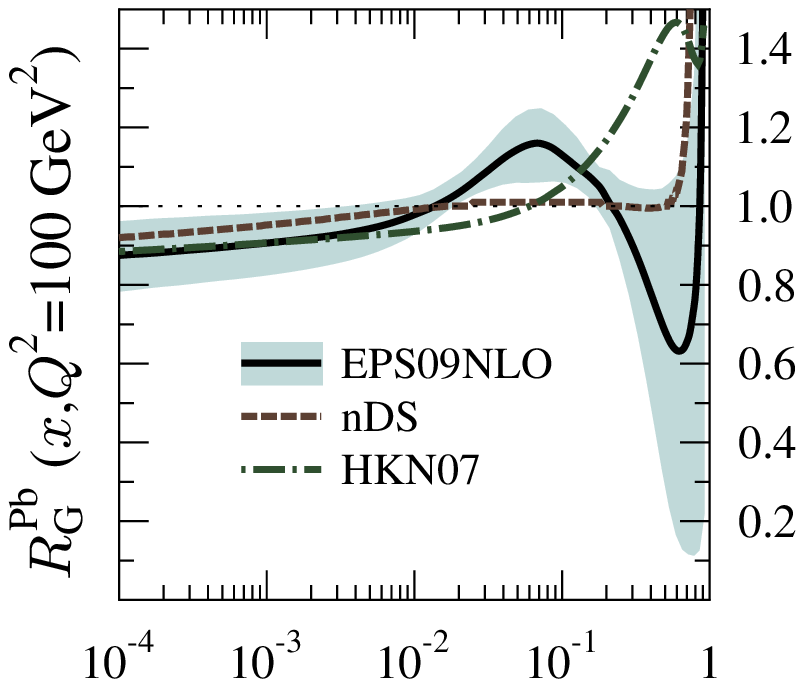}
    \end{center}
    \caption{The nuclear modifications of the gluon PDF in a Pb-nucleus at $Q^2 = 1.69$~GeV$^2$ (left) and $Q^2 = 100$~GeV$^2$ (right) in the EPS09 set (and their uncertainties), nDS set and HKN set.}
    \label{fig:RPb}
\end{figure}

The uncertainties related to these modifications are still significant especially in the low-$Q^2$ gluon sector, as shown in Fig.~\ref{fig:RPb} (left panel) by the EPS09 uncertainty band on the ratio $R_g^{\rm Pb}$ at $Q^2=1.69$~GeV$^2$.
The small-$x$ uncertainties decrease, however, towards higher scales but the large-$x$ uncertainties remain, as is shown by the right panel of Fig.~\ref{fig:RPb}. In the EPS09 set, the uncertainties of $R_i^A(x,Q^2)$ are encoded through 15 pairs of error sets, $S_{k=1,15}^\pm$, whose gluon modifications are plotted as a light blue band in
Fig.~\ref{fig:RPb}. The propagation of the nPDF uncertainties into a physical quantity $X$, such as the gluon modification in Fig.~\ref{fig:RPb} or prompt photon production considered here, can be obtained by squaring the deviations from the central result (which is obtained with the best fit, the set $S_0$) using the following prescription  \cite{Eskola:2009uj}:
\begin{eqnarray}
(\Delta X^+)^2  & \approx &  \sum_i \left[ \max\left\{ X(S^+_i)-X(S_0), X(S^-_i)-X(S_0),0 \right\} \right]^2,\nonumber\\ 
(\Delta X^-)^2  & \approx & \sum_i \left[ \max\left\{ X(S_0)-X(S^+_i), X(S_0)-X(S^-_i),0 \right\} \right]^2.
\label{eq:X_Extremum3}
\end{eqnarray}
For more details about the nPDFs and their uncertainties, see Ref.~\cite{Eskola:2009uj}.

\section{Probing nPDFs in $p$--$A$ collisions}\label{sec:pA}

\subsection{Feasibility of direct photon measurements}\label{sec:xsabsolute}

\begin{figure}[h]
    \begin{center}
      \includegraphics[height=6.4cm]{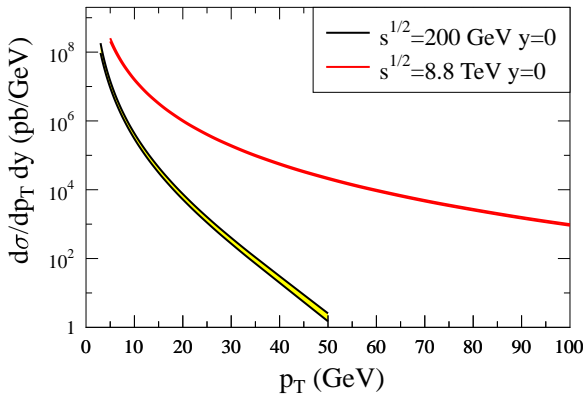}
    \end{center}
    \caption{$d\sigma/d\pt dy$ inclusive photon production $\pt$-spectra in minimum-bias $d$--Au collisions at RHIC ($\sqrtsnn=200$~GeV) and $p$--Pb collisions at LHC ($\sqrtsnn=8.8$~TeV). The band (not visible at LHC energy) corresponds to a scale variation from $\mu=M=M_F=[\pt/2;2\pt]$.}
    \label{fig:spectra}
\end{figure}

Before discussing the nPDF corrections to the prompt photon yield in nuclear collisions, we first briefly discuss the expected rates and hence the statistical accuracy at RHIC and LHC. In Fig.~\ref{fig:spectra} we plot the absolute NLO prompt photon $\pt$-spectra at mid-rapidity in (minimum-bias) $d$--Au collisions at RHIC and $p$--Pb collisions at the LHC, together with their scale uncertainty (band).  Assuming that it would be necessary to reach a $\delta = 10\%$ statistical accuracy in order to measure meaningfully the nuclear effects in prompt photon production, one would need about $\mathcal{N} = \delta^{-2} \approx 100$ events
in a ($\pt$, rapidity)-bin. 

The foreseen luminosity in $d$--Au collisions at RHIC-II is ${\cal L}^{year}=0.74$~pb$^{-1}$
assuming 12 weeks of ion runs per year~\cite{RHIC:WorkReport}. Requiring ${\cal N}=100$ events/GeV, the {\it minimal} cross section reads
\begin{equation}
\frac{d\sigma^{\rm d Au}}{d\pt dy}
= \frac{\mathcal{N}}{\mathcal{L}^{\rm int}}
 \approx \frac{100}{0.74}\  {\rm pb}/{\rm GeV} 
\approx 1.4\times10^{2} {\rm pb}/{\rm GeV}.
\end{equation}
which would be reached at $\pt \simeq 35$~GeV, see Fig.~\ref{fig:spectra}.
At LHC such a statistical accuracy would correspond to a cross section
\begin{equation}
\frac{d\sigma^{\rm p {\rm Pb}}}{d\pt dy}
 = \frac{100}{0.1}\  {\rm pb}/{\rm GeV} 
\approx 10^3\ {\rm pb}/{\rm GeV},
\end{equation}
assuming a rather conservative ${\cal L}=10^{29}~{\rm cm}^{-2}s^{-1}$ leading to a yearly integrated integrated luminosity\footnote{We assume here that the LHC will run one month per year in the ion mode, which is taken to be $\Delta t\equiv10^6$~s by convention.} $\mathcal{L}^{\rm int}=0.1$~pb$^{-1}$~\cite{Accardi:2004be}. From Fig.~\ref{fig:spectra}, this precision is achieved for transverse momenta up to $\pt\simeq100$~GeV in one year.
These considerations set the $\pt$-windows for the LHC and RHIC in which we 
perform our computations.
Better precision could only be achieved of course in a more limited $\pt$ range.  A more satisfactory $\delta=3\%$ precision would be reached for transverse momenta less than $\pt\simeq 25$~GeV at RHIC and $\pt\simeq60$~GeV at the LHC.

\subsection{RHIC}

Let us now move to the prediction of the nuclear 
modification
ratio, or ``quenching factor'', in $d$--Au collisions at RHIC,
 defined as
\begin{equation}\label{eq:quenchingfactor}
R_{d {\rm Au}}^\gamma \equiv \frac{{d\sigma/d\pt}  \left( d + {\rm Au \rightarrow \gamma + X}\right)}
                      {2\times197 \times {d\sigma/d\pt} \left( p + p \rightarrow \gamma + {\rm X}\right)}.
\end{equation}
Note that the quenching factor (\ref{eq:quenchingfactor}) is not normalized to one when no nuclear modifications in the parton densities are assumed. The reason comes from an ``isospin'' effect as the density of up quarks --~to which photons mostly couple~-- in the deuterium and the gold nucleus is smaller than that in a
proton because of the presence of neutrons inside nuclei~\cite{Arleo:2006xb}. These corrections should be most pronounced whenever the valence quark sector of the nuclei is probed, that is at large $x_1/x_2$ and thus large $\pt$. On the contrary, no 
significant isospin effect is expected in charge-averaged or neutral
hadron production which do not involve QED couplings.
 
\begin{figure}[h]
  \begin{minipage}[t]{7.1cm}
    \begin{center}
      \includegraphics[height=6.2cm]{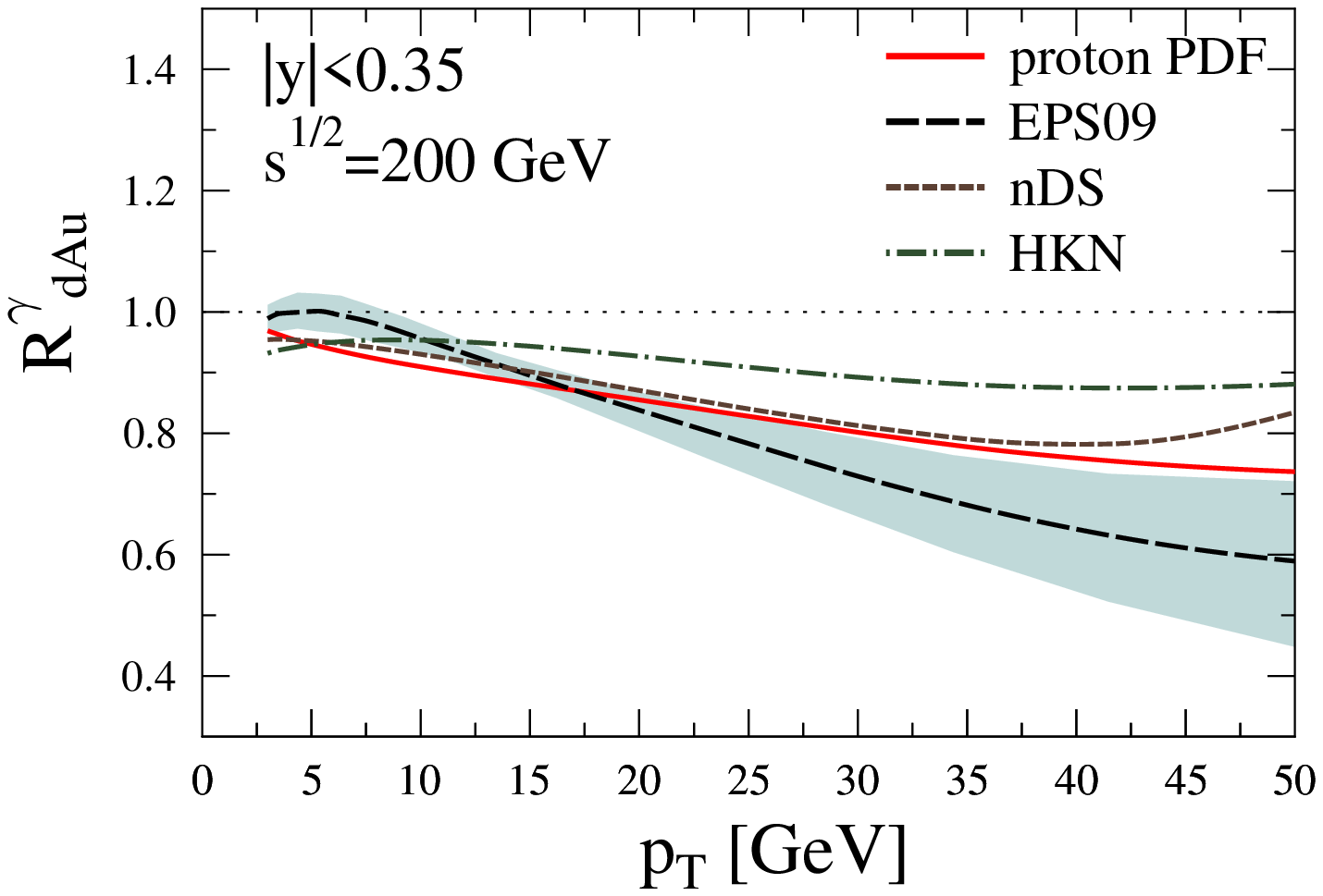}
    \end{center}
  \end{minipage}
 ~
  \begin{minipage}[t]{7.1cm}
    \begin{center}
      \includegraphics[height=6.2cm]{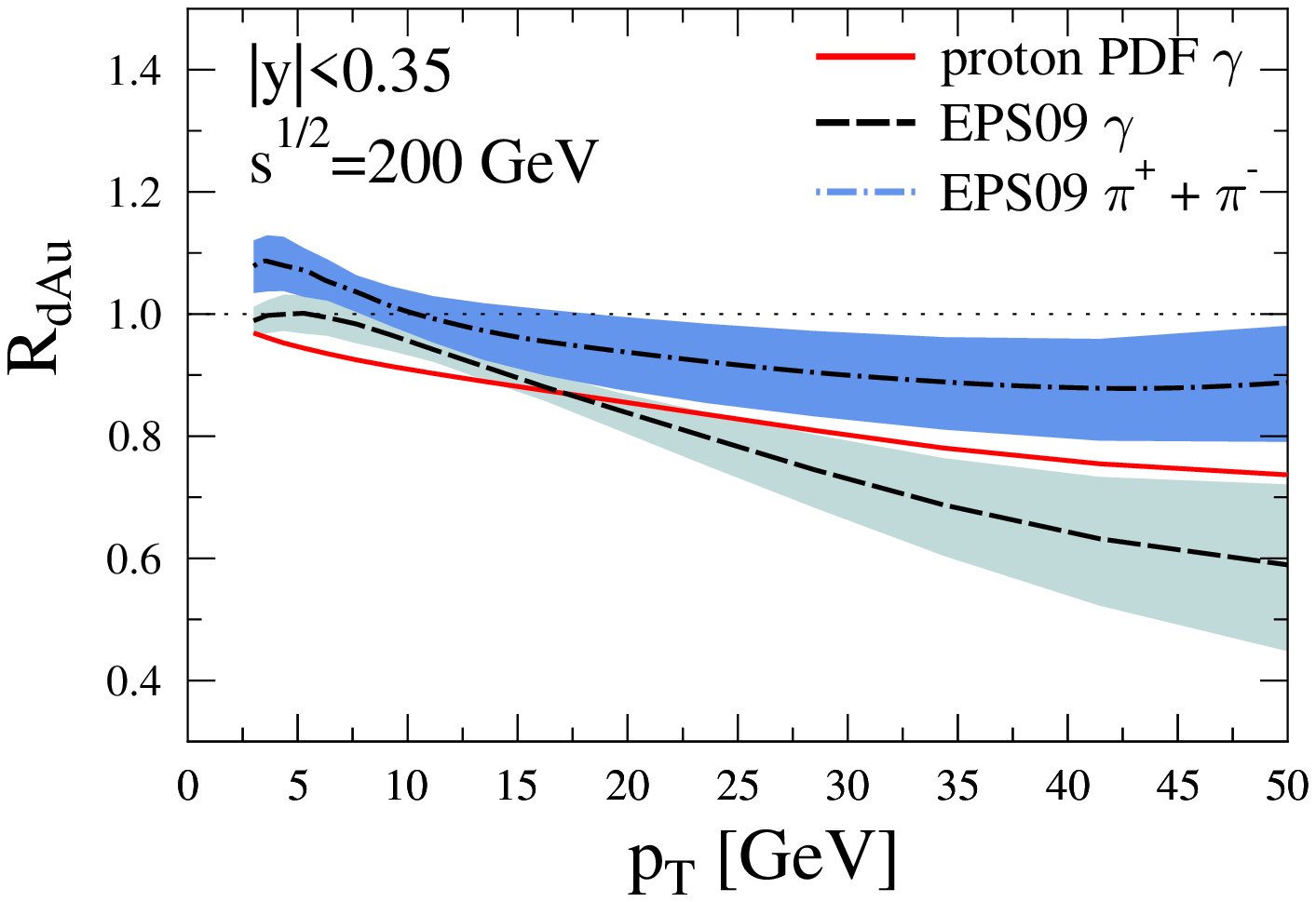}
    \end{center}
  \end{minipage}
    \caption{Nuclear modification ratio $\rdau^{\gamma}$ of inclusive photon production at $|y|\le0.35$ in $d$--Au collisions at $\sqrtsnn=200$~GeV using EPS09 nPDFs, in comparison to (i) nDS and HKN sets (left), and (ii) the pion case, $\rdau^\pi$ (right).}
    \label{fig:photon_y00_0200}
\end{figure}

The predictions for $R_{d {\rm Au}}^\gamma$ at mid-rapidity and $\sqrtsnn=200$~GeV are shown in Fig.~\ref{fig:photon_y00_0200} (left) in the transverse momentum range $\pt = 5$--50~GeV.
The central EPS09 prediction is shown as a dashed black line, the light blue band corresponding to its 
uncertainty range, while the nDS (resp. HKN) prediction is denoted by the dotted brown (resp. dash-dotted green) curve. For separating the genuine nuclear effects from the sheer isospin effects, the solid red line indicates the calculation with no 
nuclear modifications in the PDFs (i.e. just the free proton PDFs are used) -- it also serves as a comparison baseline for the nuclear effects obtained using the various nPDF sets.

The relative difference between EPS09 with respect to the baseline (red curve) prediction  follows roughly the shape of the nuclear modifications in Fig.~\ref{fig:RPb}. Below $\pt\lesssim10$~GeV photon production is sensitive to anti-shadowing corrections ($x_2\sim2\pt/\sqrtsnn\simeq0.1$) while it is suppressed at larger momenta due to the EMC effect. This is also the case for pion production for
which we only show the EPS09 prediction on $R_{d Au}^\pi$ as a blue band in the right panel of Fig.~\ref{fig:photon_y00_0200}, except that in this case the suppression is 
practically free of isospin corrections, as mentioned above.

\begin{figure}[h]
  \begin{minipage}[t]{7.1cm}
    \begin{center}
      \includegraphics[height=6.2cm]{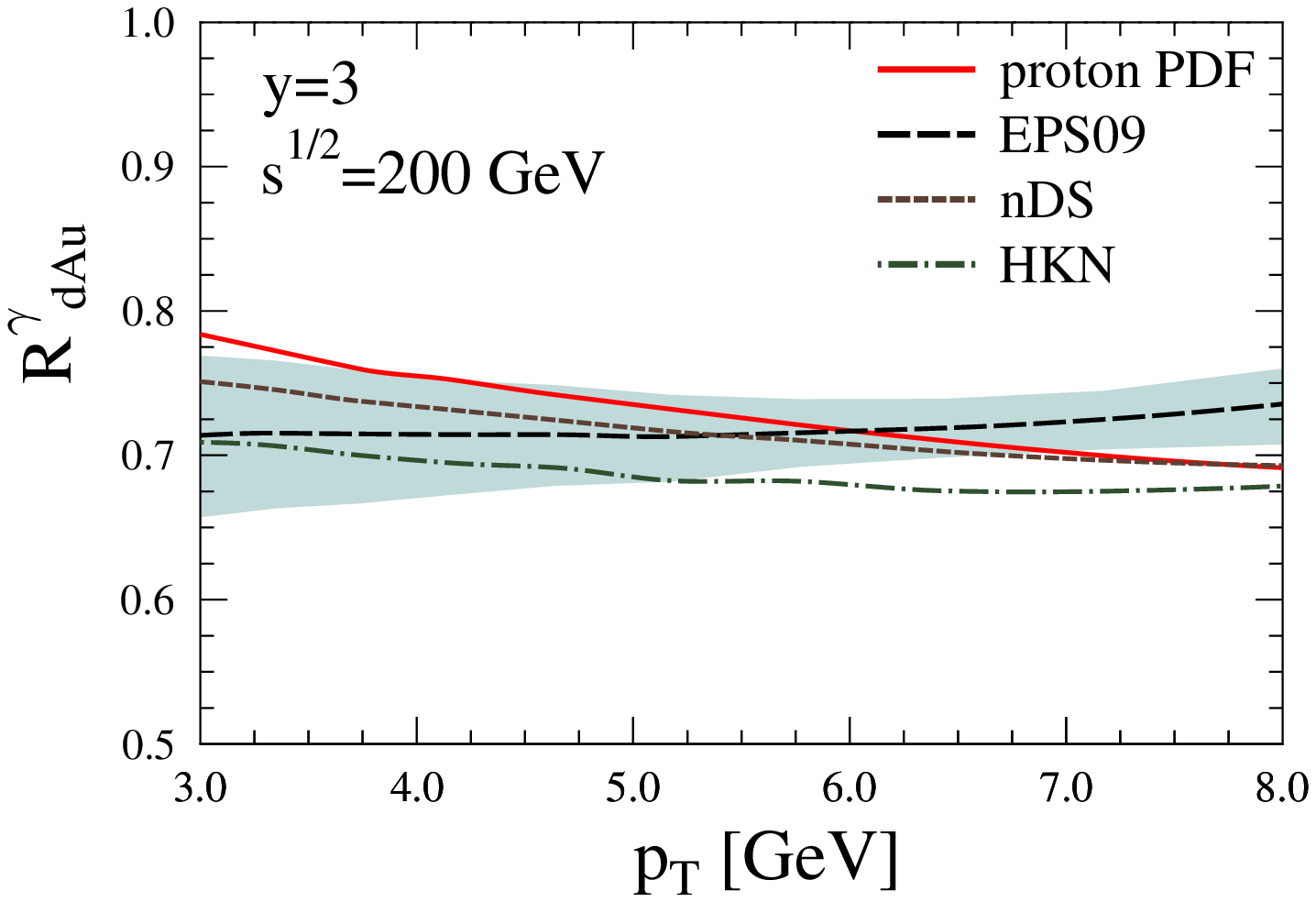}
    \end{center}
  \end{minipage}
 ~
  \begin{minipage}[t]{7.1cm}
    \begin{center}
      \includegraphics[height=6.2cm]{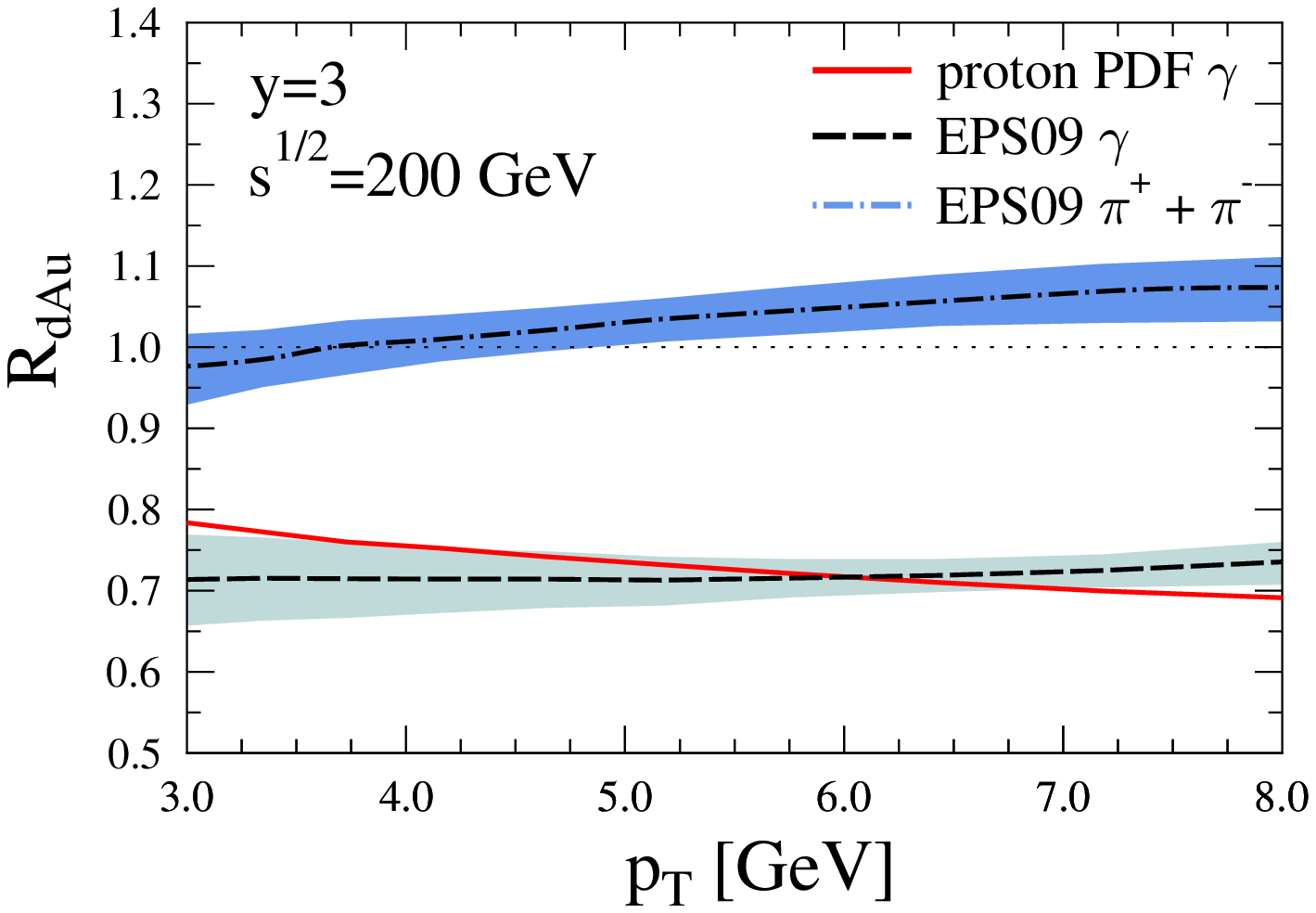}
    \end{center}
  \end{minipage}
    \caption{Same as Fig.~\ref{fig:photon_y00_0200} at $y=3$.}
    \label{fig:photon_y30_0200}
\end{figure}

 As noted in section~\ref{sec:ingredients}, prompt photon production at forward rapidities, $y>0$, is more sensitive to lower momentum fractions in a target nucleus, $x_2\propto\exp(-y)$, than the mid-rapidity production, and therefore it probes efficiently the gluon nPDFs at small $x$. We demonstrate this in Fig.~\ref{fig:photon_y30_0200} (left)
by plotting the analogue of Fig.~\ref{fig:photon_y00_0200} (left) at rapidity $y=3$ and in a narrower $\pt$-range due to phase-space restriction. As expected, $\rdau^\gamma$ at small $\pt$ falls now below the baseline 
(red) curve, signaling that this kinematical domain lies in the small-$x$ shadowing region ($\pt=5$~GeV and $y=3$ corresponds to a typical value $x_2\sim 2\pt e^{-y}/\sqrtsnn\sim 2\times10^{-3}$). From these figures, however,  one may see that the predicted differences due to nPDF effects are not very large -- of the order of 10\% or less. For completeness, the quenching factor of single-pion production, $\rdau^\pi$, at $y=3$ is also plotted together with $\rdau^\gamma$ in Fig.~\ref{fig:photon_y30_0200} (right). The behaviour of $\rdau^\pi$ is again roughly similar to 
$\rdau^\gamma$ except for the isospin corrections.

It might be surprising at first glance to observe a strong isospin effect in the prompt photon channel at forward rapidity since quark distributions in a proton and in a neutron are symmetric at {\it small} values of $x$: $u^p(x)=d^n(x)\simeq d^p(x)=u^n(x)$. As a matter of fact, the strong isospin corrections visible in Fig.~\ref{fig:photon_y30_0200} actually come from the {\it deuteron} projectile which is probed at large $x_1 \propto \exp(+y)$. As we shall see in the next section, unlike in the RHIC case, the isospin effect vanishes at forward rapidity at the LHC since the same projectile is used in $p$--$p$ and $p$--Pb collisions.

\begin{figure}[h]
  \begin{minipage}[t]{7.1cm}
    \begin{center}
      \includegraphics[height=6.2cm]{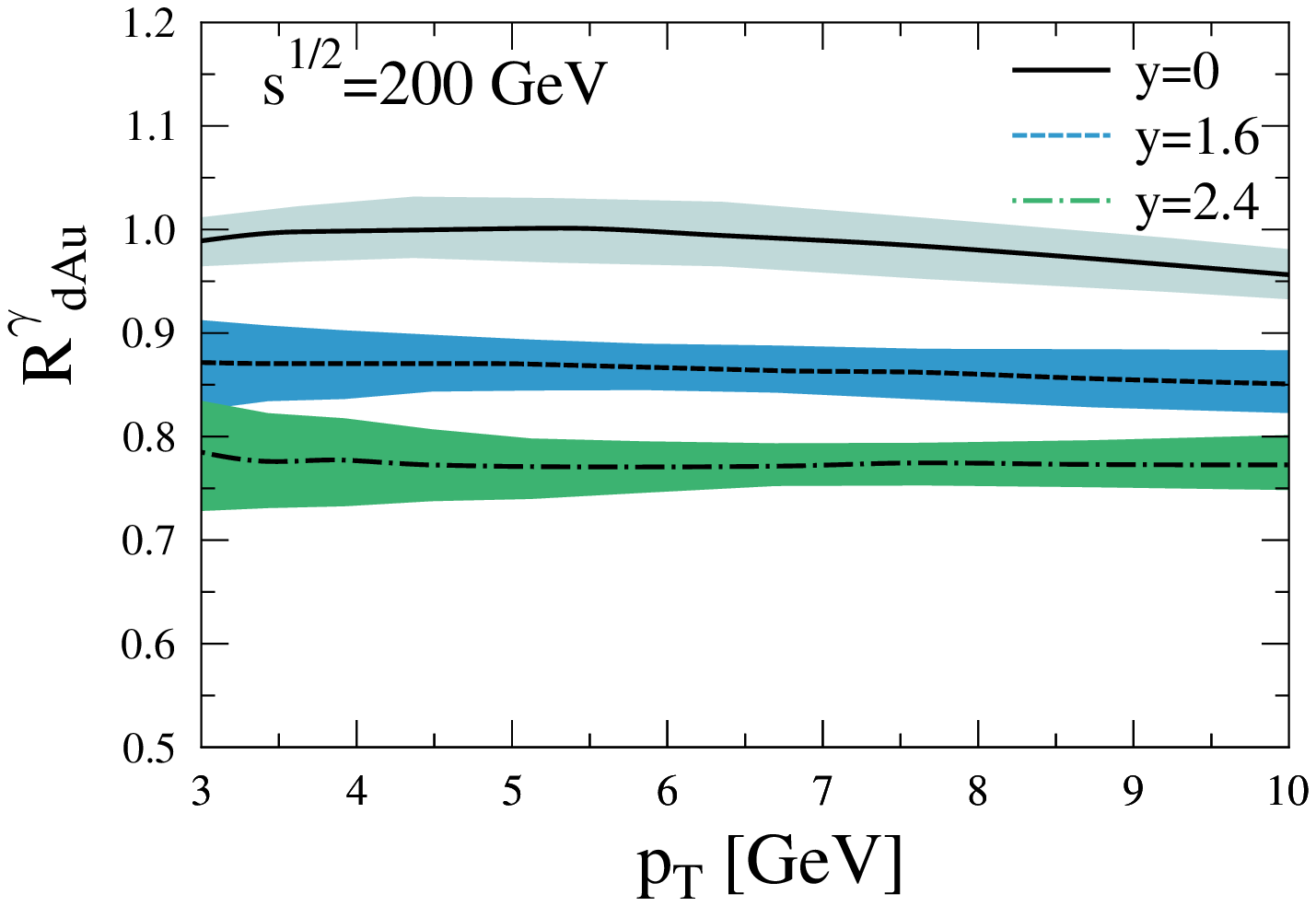}
    \end{center}
  \end{minipage}
 ~
  \begin{minipage}[t]{7.1cm}
    \begin{center}
      \includegraphics[height=6.2cm]{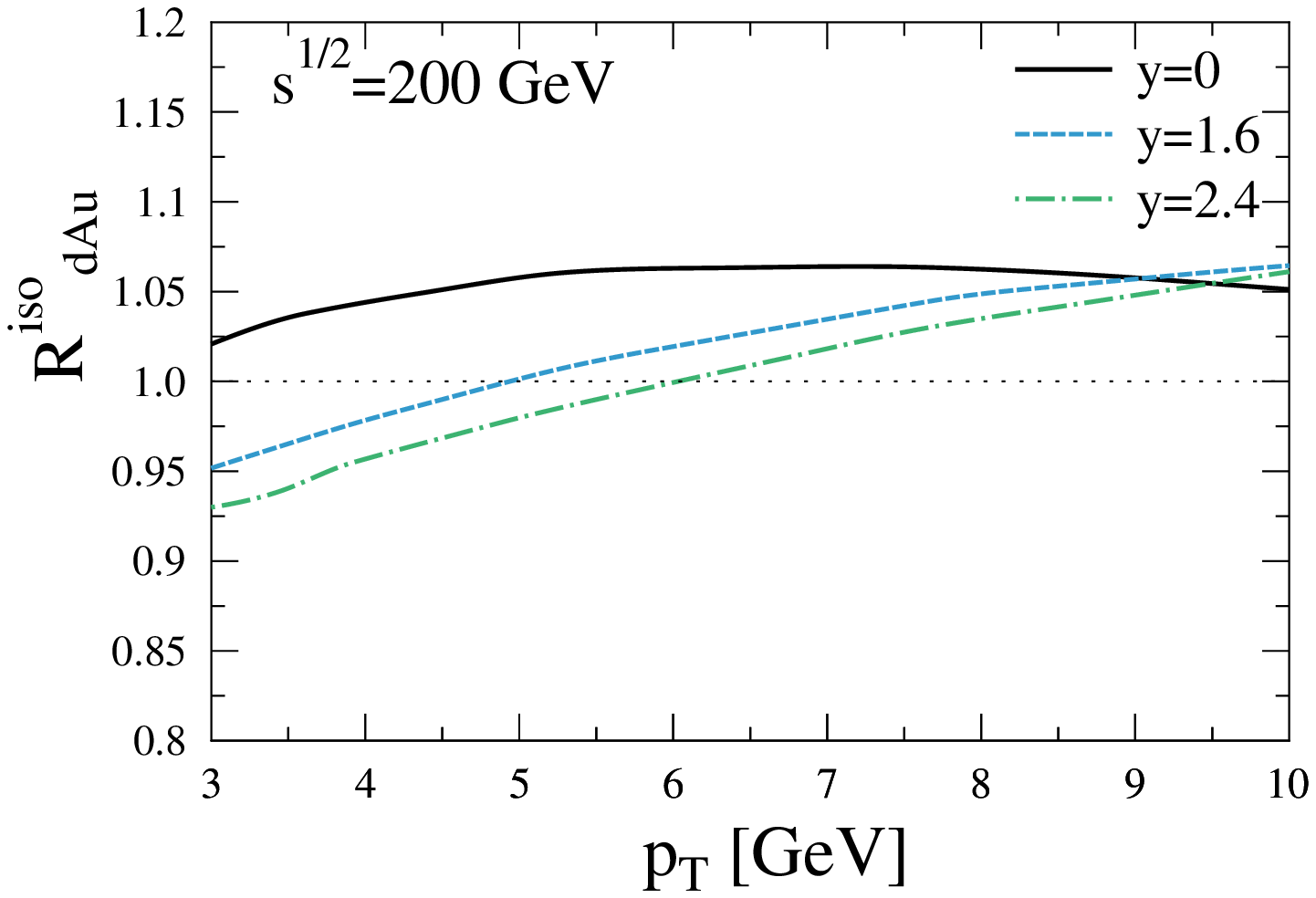}
    \end{center}
  \end{minipage}
    \caption{Left: Prompt photon quenching factor $\rdau^\gamma$ as a function of $\pt$ in different rapidities: $y=0$ (solid), $y=1.6$ (dotted) and $y=2.4$ (dash-dotted) with EPS09 nPDF corrections. Right: Double ratio $R^{\rm iso}_{_{d {\rm Au}}} \equiv\rdau^\gamma({\rm EPS09})/\rdau^\gamma({\rm proton\ PDFs})$.}
    \label{fig:photon_y00_0200_y}
\end{figure}

The RHIC photon data will be useful to learn about the behaviour of the gluon nPDFs from small to large $x$. In order to show the different kinematical regions probed at forward rapidities, $\rdau^\gamma$ is plotted in Fig.~\ref{fig:photon_y00_0200_y} (left) as a function of $\pt$ for three rapidity bins ($y=0$, $y=1.6$, and $y=2.4$) which should be accessible by the PHENIX experiment at RHIC. The predictions are obtained assuming the EPS09 nPDFs. The EPS09 $R_{dAu}^{\gamma}$ ratio looks remarkably flat at all rapidity bins, even though the normalization differs: the larger the rapidity, the stronger the shadowing 
and the smaller the quenching factor.  This flat behaviour is however somewhat accidental as it results from the interplay of isospin and nPDF effects. As can be seen in Fig.~\ref{fig:photon_y00_0200_y} (right) where the quenching factor is normalized with what is expected assuming isospin effects only, the behaviour is very different\footnote{This would correspond roughly to the production ratio $\propto d^3\sigma  \left( d + {\rm Au \rightarrow \gamma + X}\right)\ /\ d^3\sigma \left( d + d \rightarrow \gamma + {\rm X}\right)$ in which isospin corrections are very small.}. At mid-rapidity, prompt photon production is sensitive to anti-shadowing effects in this $\pt$-range (see also Fig.~\ref{fig:photon_y00_0200}) whereas the crossing region between anti-shadowing and shadowing is probed at $y=1.6$ and $y=2.4$ around $\pt=5$--6~GeV.

\subsection{LHC}

\begin{figure}[h]
  \begin{minipage}[t]{7.1cm}
    \begin{center}
      \includegraphics[height=6.2cm]{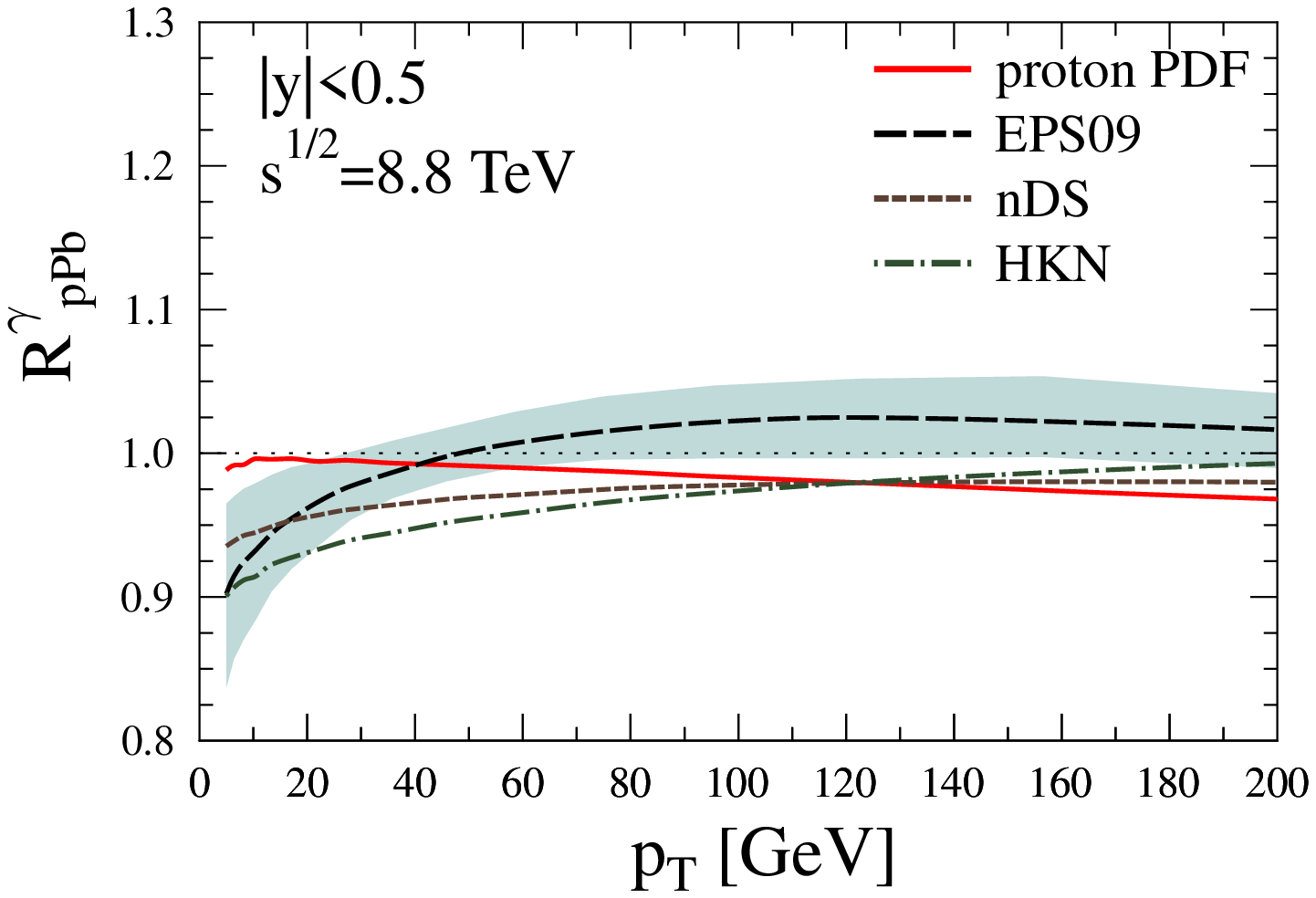}
    \end{center}
  \end{minipage}
 ~
  \begin{minipage}[t]{7.1cm}
    \begin{center}
      \includegraphics[height=6.2cm]{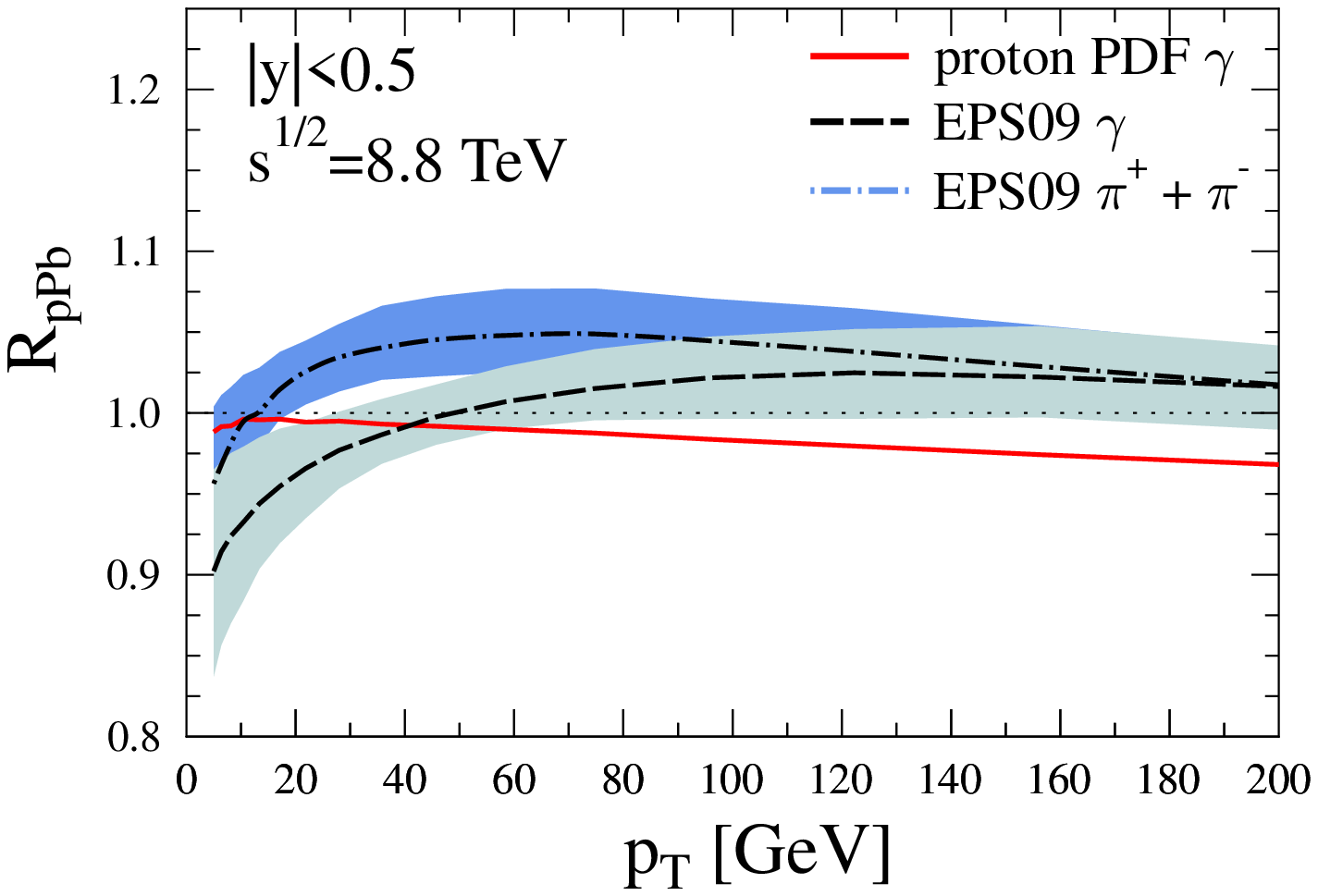}
    \end{center}
  \end{minipage}
    \caption{Nuclear modification ratio $\rdau^\gamma$ of inclusive photon production at $|y|\le0.5$ in $p$--Pb collisions at $\sqrtsnn=8.8$~TeV using the EPS09 nPDF set, in comparison (left) to nDS and HKN nPDFs and (right) to the pion production channel.}
    \label{fig:photon_y00_8800}
\end{figure}

In this section we give predictions for  $p$--Pb collisions at the LHC. The nuclear modification factor,

\begin{equation}
R_{p {\rm Pb}}^{\gamma} \equiv \frac{{d\sigma/d\pt}  \left( p + {\rm Pb \rightarrow \gamma + X}\right)}
                      {208\times{d\sigma/d\pt} \left( p + p \rightarrow \gamma + X\right)},
\end{equation}
is plotted for prompt photon production at mid-rapidity in Fig.~\ref{fig:photon_y00_8800}. Because of the larger
center-of-mass energy, $\sqrtsnn=8800$~GeV, the typical values of $x_2 \propto 1/\sqrtsnn$  probed in the nuclear target are much smaller than at RHIC. Remarkably, 
the differences in the results obtained with the different nPDF sets, are now more pronounced.

In the EPS09 set (light blue band), prompt photon production is enhanced with respect to the isospin effects (solid red line) above $\pt=20$--40~GeV due to the anti-shadowing. On the contrary, the transition from the shadowing to the anti-shadowing regions occurs only above $\pt \simeq 120$~GeV when using nDS (brown short-dashed) and HKN (green dot-dashed) nPDF sets. Like at RHIC, it is particularly interesting to note that the predictions using the latter two sets fall outside the EPS09 uncertainty band. As a consequence, this observable --~which should be easily accessible at LHC~-- will allow one to set 
tighter constraints to the nuclear gluon densities.
The EMC-effect (appearing at $x_2\gtrsim 0.1$) which can be probed easily at RHIC would become visible at the LHC only at very large $\pt$, $\pt\gtrsim 500$~GeV at $y=0$, or at negative rapidities, say, at $y<-2.2$ for $\pt=50$~GeV.

\begin{figure}[h]
  \begin{minipage}[t]{7.1cm}
    \begin{center}
      \includegraphics[height=6.2cm]{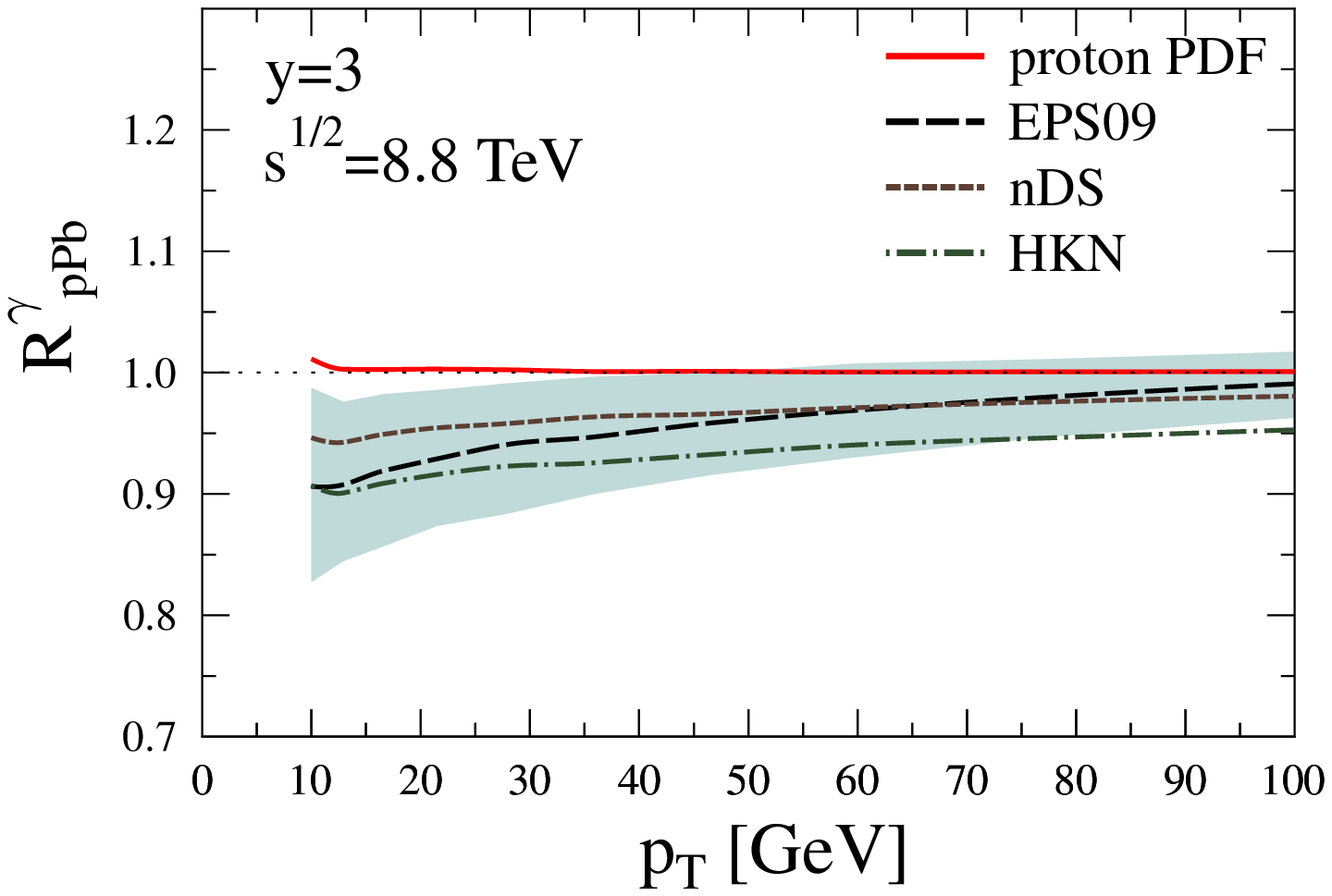}
    \end{center}
  \end{minipage}
 ~
  \begin{minipage}[t]{7.1cm}
    \begin{center}
      \includegraphics[height=6.2cm]{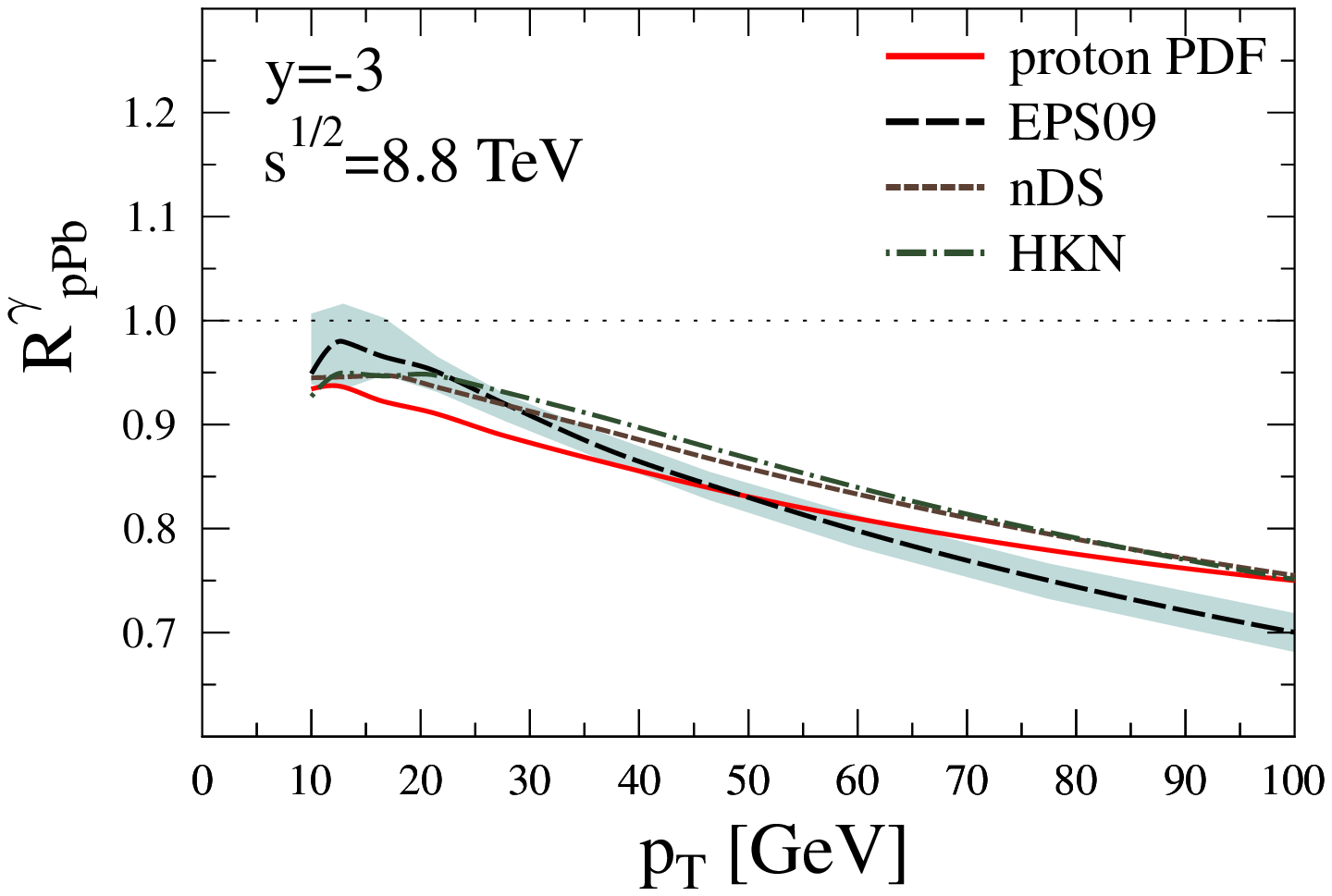}
    \end{center}
  \end{minipage}
    \caption{Nuclear production ratio $\rdau^\gamma$ of inclusive photon production at $y=3$ (left) and $y=-3$ (right) in $p$--Pb collisions at $\sqrtsnn=8.8$~TeV using the EPS09 nPDF set, in comparison to the nDS and HKN nPDF sets.}
    \label{fig:photon_y30_8800}
\end{figure}

The smallest $x_2$ values can be reached at LHC (because of the high c.m.s. energy), at forward rapidity and at not too large values of $\pt$. This kinematic region is therefore ideal for probing efficiently the shadowing region. This can be seen in Fig.~\ref{fig:photon_y30_8800} (left) where the photon suppression at $y=3$ is plotted as a function of $\pt$. As discussed in the previous section, no isospin corrections (solid red line) are expected since at small values of $x_2$ the quark distributions in protons and neutrons are identical. Without a surprise, the photon quenching factor is below 1 because of the shadowing in all the nPDF sets 
studied here. 
The large uncertainty band of the EPS09 predictions in Fig.~\ref{fig:photon_y30_8800} reflects that of $R_g^{\rm Pb}$ shown in Fig.~\ref{fig:RPb}. Therefore, high-precision measurements at $\pt\simeq10$~GeV and at $y=3$ in $p$--Pb collisions at LHC 
would bring significant further constraints for the nPDF global fit analyses.

Going to the backward direction at LHC would allow one to probe nPDFs in an $x$-region similar to that reached with mid-rapidity photon production at RHIC energy. As discussed in~\cite{Arleo:2007js}, measuring prompt photon production in $p$--$A$ collisions at negative rapidity would allow one to access the {\it quark} nPDFs at large values of $x_2$, $\rpa^{^{y<0}}(\xt) \simeq \rafd\left(\xt \exp(-y)\right)$, while mid-rapidity photon production is sensitive to {\it both} the quark and gluon nPDFs, $\rpa^{^{y=0}}(\xt) \simeq ((\rafd(\xt)+\rag(\xt))/2$. 
Fig.~\ref{fig:photon_y30_8800} (right) shows the expected photon suppression in the rapidity bin $y=-3$ in the range $\pt=10$--$100$~GeV. This range would correspond at RHIC, at similar values of $x_2$, to transverse momenta $\pt=\cO{\sqrtsnn^{\rm RHIC}/\sqrtsnn^{\rm LHC}\times e^3}=5$--$50$~GeV at mid-rapidity. As can be seen from the comparison of Fig.~\ref{fig:photon_y30_8800} (right) with Fig.~\ref{fig:photon_y00_0200} (left), the expected photon suppression is rather similar. Remarkably, the spread of the EPS09 theoretical predictions proves narrower at the LHC than at RHIC, reflecting the fact that quark nPDFs are much better constrained than the gluon nPDFs at large values of $x$~\cite{Eskola:2009uj}\footnote{In addition, larger scales are probed at the LHC, $Q^2\big|_{\rm LHC} \sim 4\times Q^2\big|_{\rm RHIC}$. This is however a rather moderate effect since the EPS09 gluon nPDF ratios do not exhibit a strong $Q^2$-dependence at large $x$~\cite{Eskola:2009uj}.}.
 
\subsection{Scale dependence}

\begin{figure}[h]
    \begin{center}
      \includegraphics[height=6.4cm]{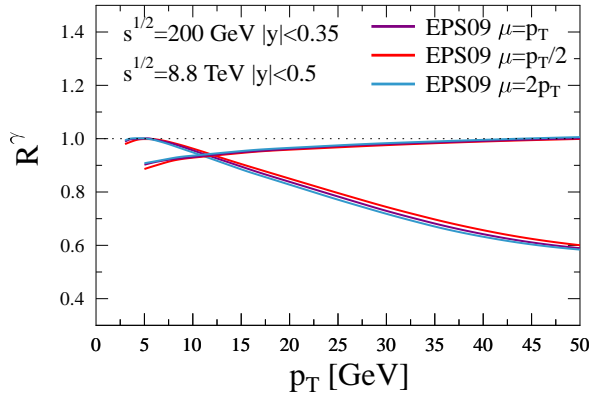}
    \end{center}
    \caption{Scale dependence of the nuclear modification ratio $\rdau^\gamma$ for mid-rapidity prompt photon production at RHIC (lower curve) and LHC (upper curve). Calculations are performed using the central EPS09 nPDF set and varying all scales by a factor of two with respect to the photon transverse momentum; see the text for details.}
    \label{fig:scaledep}
\end{figure}

As discussed in section~\ref{sec:ingredients}, collinearly factorized pQCD cross sections depend on the renormalization and factorization scales, which all are of the order of the photon $\pt$.
In the absolute cross sections the sensitivity to these scales reflects the uncertainty which results from terminating the perturbation series at a certain order, and thus neglecting the higher-order corrections (here, NNLO and beyond).
In the nuclear modification ratio $\rdau^\gamma$, such scale uncertainties should nevertheless largely just
cancel out. However, since the nPDF corrections $R_i^{A}(x, M^2)$ (Eq.~(\ref{eq:npdfratio})) do depend on the factorization scale $M$, also $\rdau^\gamma$ may exhibit some dependence on $M$. To quantify this theoretical uncertainty, the nuclear modification ratio of prompt photon production at mid-rapidity has been computed using the EPS09 set and varying all scales, $\mu=M=M_F$ from $\pt/2$ to $2\pt$, as was done also in the calculation of the absolute cross section in section~\ref{sec:xsabsolute}. As shown in Fig.~\ref{fig:scaledep} at the RHIC and LHC energies, the predictions show very little scale dependence. More importantly, the scale dependence proves much smaller than the 
current uncertainties in the nuclear modifications of the PDFs; see the band in Fig.~\ref{fig:spectra}.

\section{Baseline pQCD predictions in heavy-ion collisions}\label{sec:AA}

\subsection{Inclusive prompt photon production}

As can be seen in Eqs.~(\ref{eq:factorization1}) and (\ref{eq:factorization2}), nuclear corrections are expected to be much more pronounced in nucleus--nucleus collisions than in proton--nucleus collisions, 
very roughly
$R_{AA} \sim R_{pA}^2 \sim \left(R_g^A\right)^2$ in minimum-bias collisions.
Therefore, it is particularly useful to investigate how prompt photon production in heavy-ion collisions 
would show the presence of a strongly interacting QCD-medium and possibly also
constrain nuclear parton densities. This would also allow for more reliable estimates of other hard processes like jet and large-$\pt$ hadron production, which, on the basis of RHIC data (see e.g.~\cite{dEnterria:2009am} for a review) and the first LHC measurements~\cite{Aamodt:2010jd,Collaboration:2010bu}, are generally believed to
be affected by energy losses of hard partons in the formed quark-gluon plasma (QGP).  
Another motivation to make the baseline pQCD predictions for the $A$--$A$ collisions here is that the Pb--Pb programme at the LHC has already started.

\begin{figure}[h]
  \begin{minipage}[t]{7.1cm}
    \begin{center}
      \includegraphics[height=6.2cm]{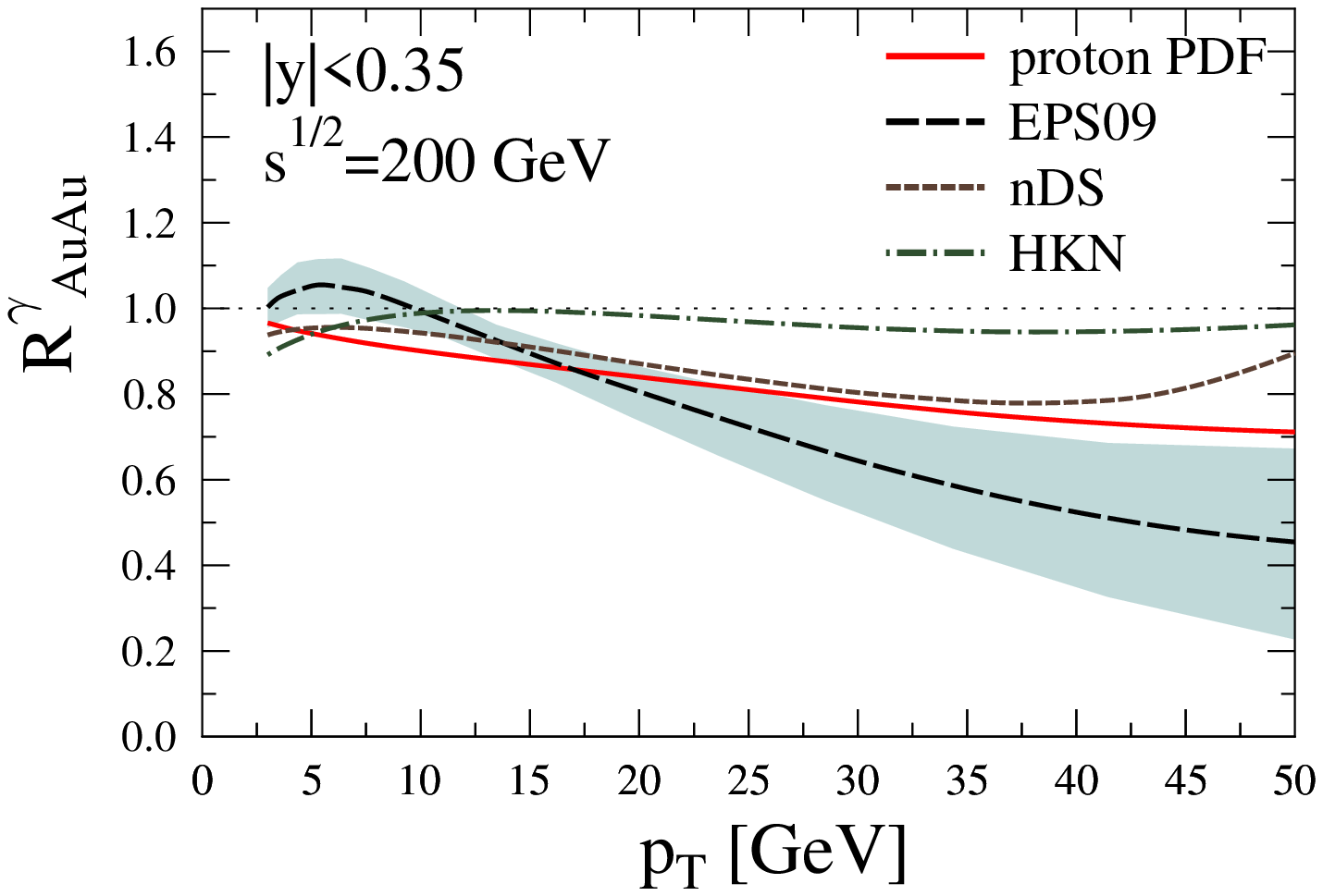}
    \end{center}
  \end{minipage}
 ~
  \begin{minipage}[t]{7.1cm}
    \begin{center}
      \includegraphics[height=6.2cm]{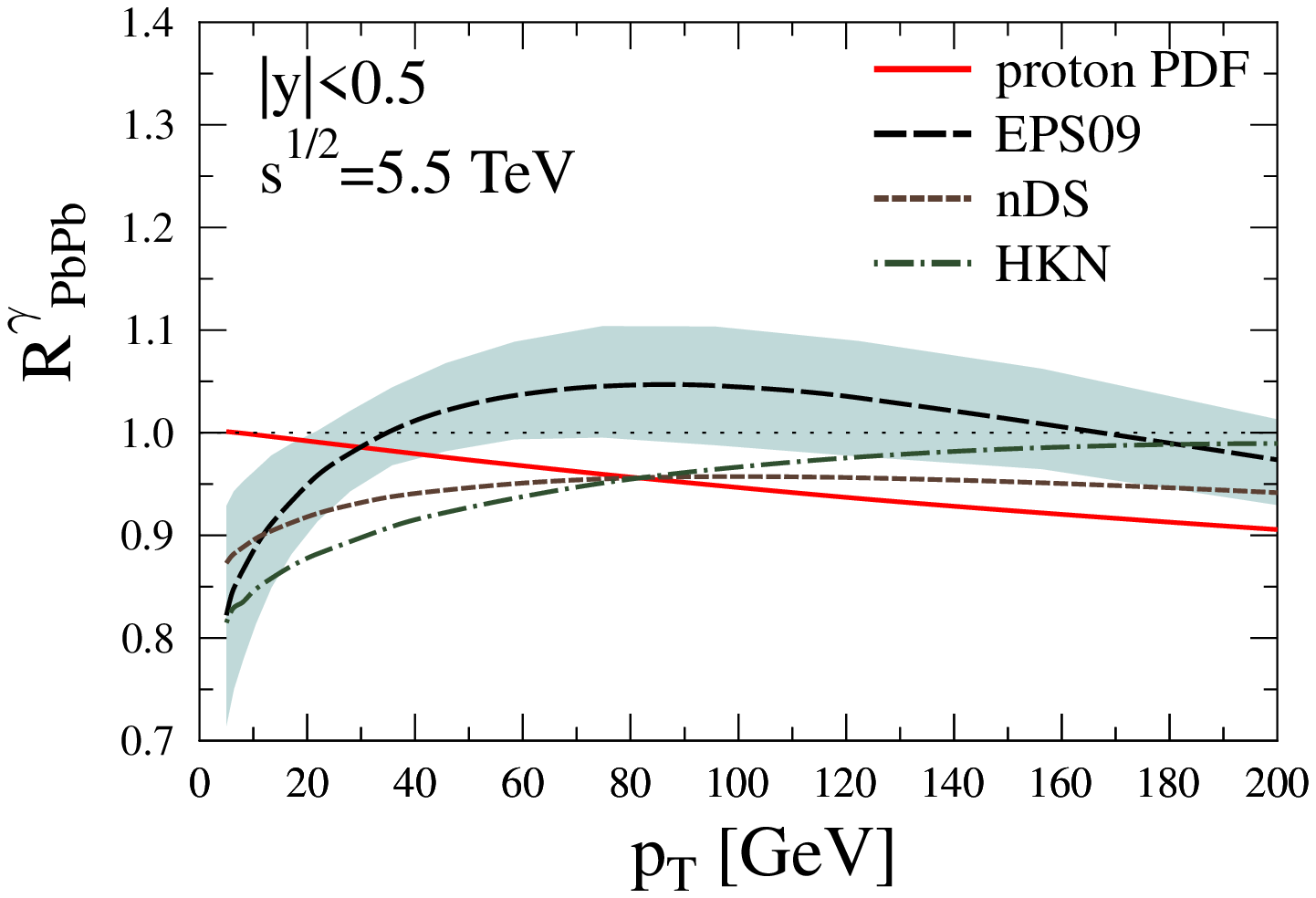}
    \end{center}
  \end{minipage}
    \caption{Left: Nuclear production ratio $\rauau^\gamma$ of inclusive photon production in minimum-bias Au--Au collisions at $\sqrtsnn=200$~GeV ($|y|\le0.35$) using EPS09, nDS and HKN nuclear parton densities. Right: same in Pb--Pb collisions at $\sqrtsnn=5.5$~TeV ($|y|\le0.5$).}
    \label{fig:photon_inclusive_AA}
\end{figure}

The quenching factors in Au--Au collisions at RHIC (left) and Pb--Pb collisions at the LHC (right) are shown in Fig.~\ref{fig:photon_inclusive_AA}. As expected, the shapes are similar but the nuclear effects are stronger than in the respective $p$--$A$ collisions, as the comparison with Fig.~\ref{fig:photon_y00_0200} (left) and Fig.~\ref{fig:photon_y00_8800} (left) shows. Note moreover that the nPDF corrections might be even more pronounced in central collisions as compared to these minimum-bias predictions, which are averaged over all centralities.
A study of the centrality dependence of $\rauau^\gamma$ would require a modeling of the spatial dependences of the nPDFs, which is outside the scope of the present study. The PHENIX collaboration at RHIC has reported on high-$\pt$ measurements of $\rauau^\gamma$ in central collisions~\cite{Isobe:2007ku} (see Fig.~2 there).  However, since these data are still preliminary and since our present setup is consistent only with minimum-bias collisions, we do not make a detailed comparison with these data in this paper.

The plots in Fig.~\ref{fig:photon_inclusive_AA} thus set the stage for the conclusions to be made from the forthcoming data: agreement with the pQCD baseline in $p$--$A$ collisions but not in $A$--$A$ collisions would be a clear hint that photons are actually sensitive to hot medium effects.

\subsection{Quenching of the fragmentation component}
\label{sec:quenching}

It is not clear yet to what extent
the production of prompt photons is affected by the presence of the dense medium in heavy-ion collisions. On the one hand, it has been suggested that the parton multiple scattering might enhance the photon yield either through jet-to-photon conversion~\cite{Fries:2002kt} or medium-induced photon bremsstrahlung~\cite{Zakharov:2004bi}. On the other hand, photon production might be suppressed due to the quenching of the fragmentation component as it happens for inclusive hadron production~\cite{Arleo:2006xb}. 

\begin{figure}[h]
  \begin{minipage}[t]{7.1cm}
    \begin{center}
      \includegraphics[height=6.2cm]{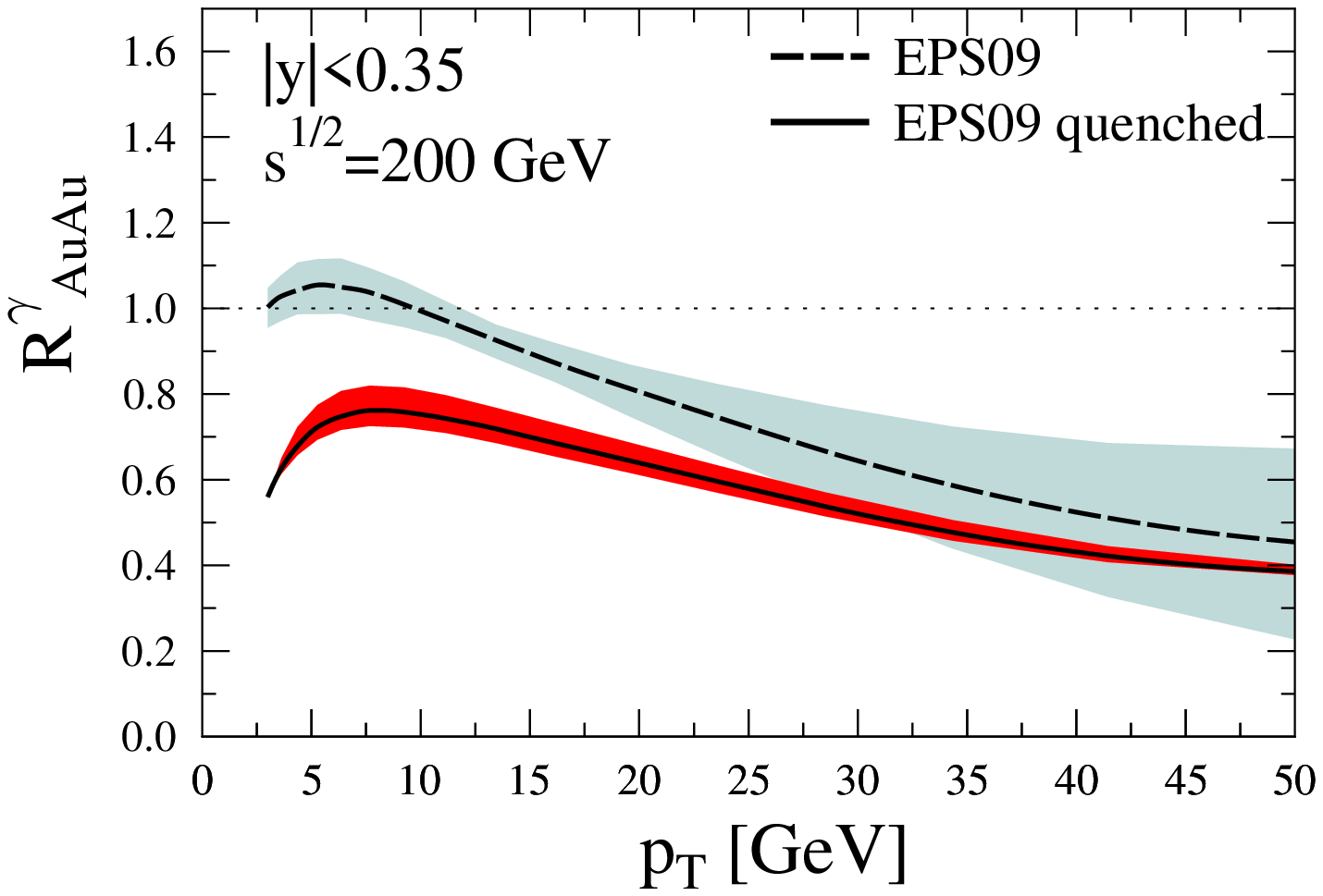}
    \end{center}
  \end{minipage}
 ~
  \begin{minipage}[t]{7.1cm}
    \begin{center}
      \includegraphics[height=6.2cm]{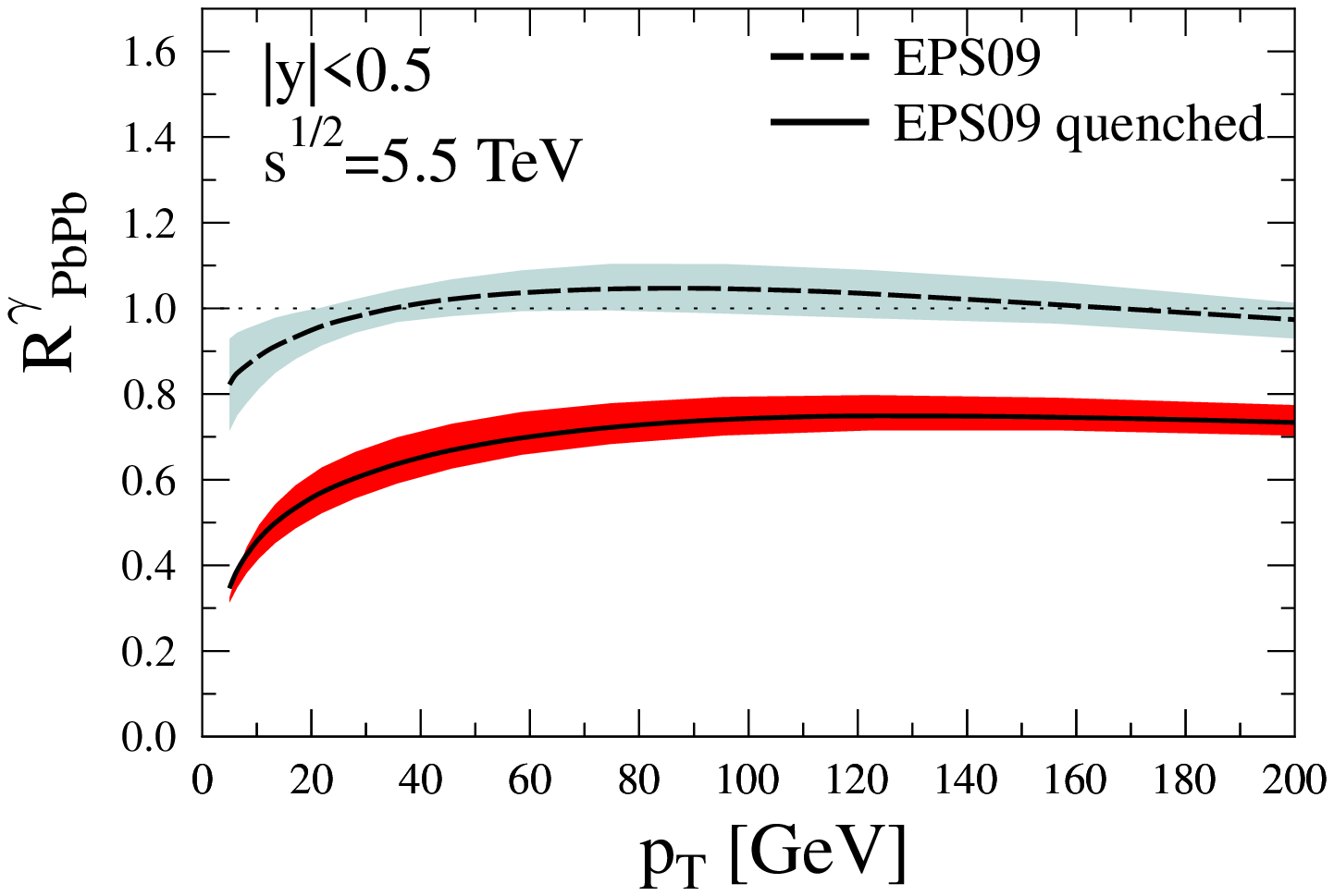}
    \end{center}
  \end{minipage}
    \caption{Upper bands: Same as Fig.~\ref{fig:photon_inclusive_AA} using the EPS09 nPDF sets. Lower bands: Nuclear production ratio $R_{AA}^\gamma$ using the central EPS09 nPDF set and after downscaling the fragmentation component by a factor of five. The band corresponds to a scale variation from $\pt/2$ to $2\pt$.}
    \label{fig:photon_quenched_AA}
\end{figure}

At large enough $\pt$, say at $\pt\gtrsim 5$~GeV at RHIC and possibly slightly above at LHC, one could nevertheless expect the suppression due to the quenching of the fragmentation component to be larger than the possible enhancement(s) caused by the medium. Therefore, we present in Fig.~\ref{fig:photon_quenched_AA} a set of curves where we have \textit{by hand} simply downscaled the fragmentation component by a factor of five (assuming in the calculation the central EPS09 set, for the illustration) which very roughly corresponds to the observed maximum suppression of  
large-$\pt$ hadrons
at RHIC and LHC. These curves should therefore serve as a lower limit for prompt photon production in $A$--$A$ collisions. It should, however, be noted that these additional curves are, strictly speaking, not physical quantities since the separation between direct and fragmentation components is ambiguous beyond the Born level and scale-dependent. As before, the calculations are performed for different scale choices from $\pt/2$ to $2\pt$ corresponding to the band in Fig.~\ref{fig:photon_quenched_AA}. We also reproduce in this figure the quenching factor obtained with the EPS09 nPDF sets (with its EPS09 uncertainty band) as shown in Fig.~\ref{fig:photon_inclusive_AA}, i.e. without any rescaling of fragmentation component, for comparison.

The prompt photon suppression is most pronounced at small $\pt$, where the fragmentation component is the largest. As $\pt$ gets larger, the direct photon contribution becomes more important, and therefore the expected photon quenching weakens. At RHIC, the photon quenching becomes comparable to the prediction with no energy loss effects above $\pt\gtrsim25$~GeV. The RHIC predictions are qualitatively similar to the calculation performed in~\cite{Arleo:2006xb}. At LHC, the direct photon contribution to the total cross section is somewhat less than at RHIC. As a consequence, energy loss effects are more pronounced and might be visible in the entire $\pt$-range.

\section{Summary}

Prompt photon production in $p$--$A$ and $d$--$A$ collisions is among the best observables for probing parton distributions in nuclei, 
especially the nuclear gluon distributions,
which are presently still poorly known.
In this paper, we have computed the nuclear modification factors of single inclusive prompt photon production at NLO accuracy using the recent nPDF sets EPS09, nDS,  and HKN. Calculations were carried out for $d$--Au collisions at RHIC and $p$--Pb collisions at the LHC, for their nominal energies. The results are conveniently complementary: 
At mid-rapidities, prompt photon production at RHIC is mostly sensitive to the gluon anti-shadowing and the EMC effect, while at the LHC these photons probe the small-$x$ shadowing region. Complementary information on the gluon shadowing region can be obtained also from the forward-rapidity prompt photons both at RHIC and the LHC.
Remarkably, by combining the future data at both colliders, one should be able to discriminate among the different existing sets of nPDFs.

On top of constraining the nuclear PDFs, inclusive prompt photon production can also be used as a ``calibration measurement'' in $A$--$A$ collisions where QGP formation is expected to quench the production of large-$\pt$ partons,
and thereby also the fragmentation component of prompt photon production. We have reported two pQCD baselines for this purpose: one by including the nuclear effects to the PDFs but no medium-effects to prompt photon production, and one by using the nPDFs and suppressing the fragmentation component by a factor of five. 

The photons produced by medium-induced bremsstrahlung or through fragmentation processes are likely to be produced alongside a significant hadronic activity in their vicinity. This is also true -- perhaps to a lesser extent -- with the jet-photon conversion process, since the initial parton will radiate before ``converting'' into a photon though a rescattering in the QGP. Therefore, a way to reduce their contributions --~and therefore medium effects~-- would be to trigger on {\it isolated} photons, i.e. using isolation criteria around each photon candidate.
In heavy-ion collisions, the high-multiplicity would make this measurement extremely challenging, hopefully made possible using recent advances developed towards the reconstruction of jets in heavy-ion collisions~\cite{Cacciari:2010te}. Predictions of isolated photon production in nuclear collisions go beyond the scope of the present paper and will be presented elsewhere.

\section*{Acknowledgements}

FA acknowledges the hospitality of the University of Jyv\"askyl\"a where this work was initiated. KJE thanks Richard Seto from PHENIX for useful discussions and the Academy of Finland, Project nr. 133005, for financial support. This work is supported by Ministerio de Ciencia e Innovaci\'on of Spain under
project FPA2009-06867-E; by Xunta de Galicia (Conseller\'\i a de Educaci\'on  and Conseller\'\i a de Innovaci\'on e 
Industria --- Programa Incite); by the Spanish Consolider-Ingenio
2010 Programme CPAN (CSD2007-00042); and by the European Commission grant
PERG02-GA-2007-224770. CAS is a Ram\'on y Cajal researcher.

\providecommand{\href}[2]{#2}\begingroup\raggedright\endgroup


\begin{thebibliography}{10}

\bibitem{Aurenche:1988vi}
P.~Aurenche, R.~Baier, M.~Fontannaz, J.~F. Owens and M.~Werlen,  Phys. Rev.
  {\bf D39} (1989) 3275.

\bibitem{Martin:1998sq}
A.~D. Martin, R.~G. Roberts, W.~J. Stirling and R.~S. Thorne,  Eur. Phys. J.
  {\bf C4} (1998) 463--496.

\bibitem{Aurenche:2006vj}
P.~Aurenche, M.~Fontannaz, J.-P. Guillet, E.~Pilon and M.~Werlen,  Phys. Rev.
  {\bf D73} (2006) 094007.

\bibitem{Ichou:2010wc}
R.~Ichou and D.~d'Enterria,  Phys. Rev. {\bf D82} (2010) 014015.

\bibitem{Collaboration:2010sp}
{ATLAS Collaboration},  \href{http://arXiv.org/abs/1012.4389}{{\tt 1012.4389}}.

\bibitem{Collaboration:2010fm}
{CMS Collaboration},  \href{http://arXiv.org/abs/1012.0799}{{\tt 1012.0799}}.

\bibitem{Eskola:1998df}
K.~J. Eskola, V.~J. Kolhinen and C.~A. Salgado,  Eur. Phys. J. {\bf C9} (1999)
  61--68.

\bibitem{Hirai:2004wq}
M.~Hirai, S.~Kumano and T.~H. Nagai,  Phys. Rev. {\bf C70} (2004) 044905.

\bibitem{Hirai:2001np}
M.~Hirai, S.~Kumano and M.~Miyama,  Phys. Rev. {\bf D64} (2001) 034003.

\bibitem{deFlorian:2003qf}
D.~de~Florian and R.~Sassot,  Phys. Rev. {\bf D69} (2004) 074028.

\bibitem{Hirai:2007sx}
M.~Hirai, S.~Kumano and T.~H. Nagai,  Phys. Rev. {\bf C76} (2007) 065207.

\bibitem{Eskola:2008ca}
K.~J. Eskola, H.~Paukkunen and C.~A. Salgado,  JHEP {\bf 07} (2008) 102.

\bibitem{Eskola:2009uj}
K.~J. Eskola, H.~Paukkunen and C.~A. Salgado,  JHEP {\bf 04} (2009) 065.

\bibitem{Schienbein:2009kk}
I.~Schienbein {\it et~al.},  Phys. Rev. {\bf D80} (2009) 094004.

\bibitem{Paukkunen:2010hb}
H.~Paukkunen and C.~A. Salgado,  JHEP {\bf 07} (2010) 032.

\bibitem{Kovarik:2010uv}
K.~Kovarik {\it et~al.},  \href{http://arXiv.org/abs/1012.0286}{{\tt
  1012.0286}}.

\bibitem{Arleo:2007js}
F.~Arleo and T.~Gousset,  Phys. Lett. {\bf B660} (2008) 181.

\bibitem{BrennerMariotto:2008st}
C.~Brenner Mariotto and V.~P.~Gon\c{c}alves, Phys. Rev. {\bf C78} (2008) 037901.

\bibitem{QuirogaArias:2010wh}
P.~Quiroga-Arias, J.~G. Milhano and U.~A. Wiedemann,  Phys. Rev. {\bf C82}
  (2010) 034903.

\bibitem{Stavreva:2010mw}
T.~Stavreva {\it et~al.},  JHEP {\bf 01} (2011) 152.

\bibitem{Aurenche:1987fs}
P.~Aurenche, R.~Baier, M.~Fontannaz and D.~Schiff,  Nucl. Phys. {\bf B297}
  (1988) 661.

\bibitem{Owens:1986mp}
J.~F. Owens,  Rev. Mod. Phys. {\bf 59} (1987) 465.

\bibitem{Aurenche:1998gv}
P.~Aurenche {\it et~al.},  Eur. Phys. J. {\bf C9} (1999) 107--119.

\bibitem{Aurenche:1999nz}
P.~Aurenche, M.~Fontannaz, J.-P. Guillet, B.~A. Kniehl and M.~Werlen,  Eur.
  Phys. J. {\bf C13} (2000) 347--355.

\bibitem{Nadolsky:2008zw}
P.~M. Nadolsky {\it et~al.},  Phys. Rev. {\bf D78} (2008) 013004.

\bibitem{Bourhis:1997yu}
L.~Bourhis, M.~Fontannaz and J.-P. Guillet,  Eur. Phys. J. {\bf C2} (1998) 529.

\bibitem{Albino:2008fy}
S.~Albino, B.~A. Kniehl and G.~Kramer,  Nucl. Phys. {\bf B803} (2008) 42--104.

\bibitem{RHIC:WorkReport}
A.~D. Frawley, F.~Karsch, T.~Ullrich and R.~Vogt,
  http://rhicii-science.bnl.gov/heavy/ (2006).

\bibitem{Accardi:2004be}
A.~Accardi {\it et~al.},  \href{http://arXiv.org/abs/hep-ph/0308248}{{\tt
  hep-ph/0308248}}.

\bibitem{Arleo:2006xb}
F.~Arleo,  JHEP {\bf 09} (2006) 015.

\bibitem{dEnterria:2009am}
D.~d'Enterria,  Springer Verlag. Landolt-Boernstein Vol. 1-23A, 0902.2011
  (2009).

\bibitem{Aamodt:2010jd}
{\bf ALICE}, K.~Aamodt {\it et~al.},
  \href{http://arXiv.org/abs/1012.1004}{{\tt 1012.1004}}.

\bibitem{Collaboration:2010bu}
{ATLAS Collaboration},  Phys. Rev. Lett. {\bf 105} (2010) 252303.

\bibitem{Isobe:2007ku}
{PHENIX Collaboration}, T.~Isobe {\it et~al.}, J. Phys. G {\bf 34} (2007) S1015
  [\href{http://arXiv.org/abs/nucl-ex/0701040}{{\tt nucl-ex/0701040}}].

\bibitem{Fries:2002kt}
R.~J. Fries, B.~M{\"u}ller and D.~K. Srivastava,  Phys. Rev. Lett. {\bf 90}
  (2003) 132301.

\bibitem{Zakharov:2004bi}
B.~G. Zakharov,  JETP Lett. {\bf 80} (2004) 1.

\bibitem{Cacciari:2010te}
M.~Cacciari, J.~Rojo, G.~P. Salam and G.~Soyez,  Eur. Phys. J. {\bf C71} (2011)
  1539.

\end{thebibliography}
\end{document}